\documentclass[pr,aps,superscriptaddress,floatfix,twocolumn]{revtex4}
\usepackage{graphicx}
\graphicspath{{figures/}}

\usepackage{units,xspace}

\newcommand{\micron}{\ensuremath{\unit{\mu m}}\xspace}

\usepackage{amsmath,amssymb}

\renewcommand{\vec}[1]{\boldsymbol{#1}}
\newcommand{\abs}[1]{\left\vert #1 \right\vert}

\begin{document}

\title{Colloidal transport through optical tweezer arrays}

\author{Yael Roichman}
\affiliation{Department of Physics and Center for Soft Matter Research,
New York University, New York, NY 10003}

\author{Victor Wong}
\affiliation{Stuyvesant High School, New York, NY 10282}

\author{David G. Grier}
\affiliation{Department of Physics and Center for Soft Matter Research,
New York University, New York, NY 10003}

\date{\today}

\begin{abstract}
Viscously damped particles driven past an evenly spaced array of 
potential energy wells or barriers may become kinetically
locked in to the array, or else may escape from the array.
The transition between locked-in and free-running states has
been predicted to depend sensitively on the ratio
between the particles' size and the separation between wells.
This prediction is confirmed by measurements on monodisperse
colloidal spheres driven through arrays of holographic optical traps.
\end{abstract}


\maketitle

Particles rolling down a sinusoidally modulated slope constitute
an archetype for problems as diverse as
the electrodiffusion of atoms on crystals,
the transport characteristics of Josephson junctions and the
entire field of chemical kinetics.
The one-dimensional 
tilted washboard problem has been studied exhaustively.
Considerably less is known for related problems in higher dimensions.
Recently, attention has become focused on the transport of
viscously damped colloidal particles flowing through two-dimensional
potential energy landscapes.
This system can be realized in practice
by passing fluid-borne objects through microfabricated
arrays of posts \cite{carlson97,duke97,chou99,chou00},
over arrays of electrodes 
\cite{hunt04,chiou05}
and through periodically 
structured light fields \cite{korda02b,macdonald03,ladavac04}.

\begin{figure}[!b]
  \centering
  \includegraphics[width=\columnwidth]{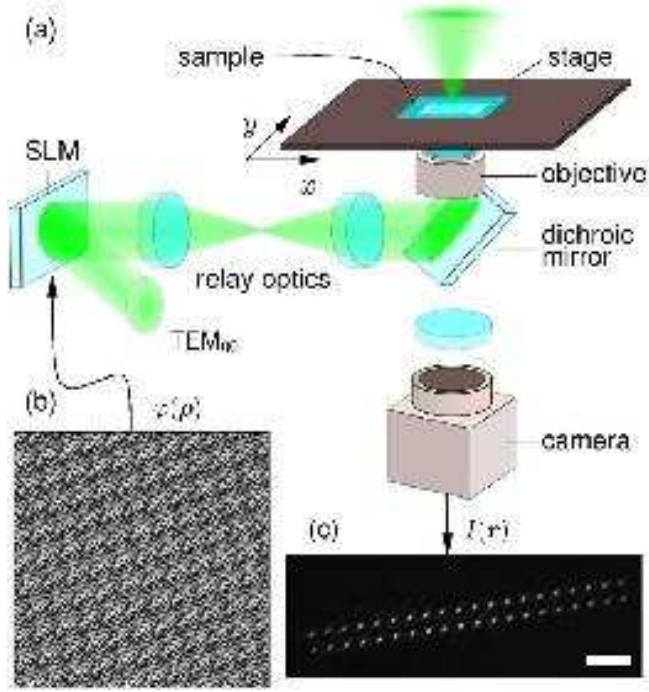}
  \caption{Colloidal transport through an array of holographic optical
  traps.  (a) Holographic trapping system.  (b) Phase hologram 
  $\varphi(\vec{\rho})$ encoding at $20 \times 2$ array of optical
  traps, shown in (c).  Scale bar indicates 5~\micron. 
 }
  \label{fig:schematic}
\end{figure}

Experimental realizations of two-dimensionally modulated transport
are interesting both because they provide insights
into the underlying fundamental problem, and also because
they constitute an entirely new category of sorting techniques.
Different types of objects, it turns out, can follow dramatically
different paths through the same physical landscape.
Sorting fluid-borne objects
by size, shape and composition have been demonstrated 
in this way \cite{macdonald03,ladavac04}.
Preliminary theoretical studies \cite{ladavac04,pelton04a}
suggest that
periodic landscapes can act as extraordinarily 
selective sieves, 
for example sorting spheres by size with exponential resolution.
This article presents experimental confirmation
of some of these theories' predictions.

Our system, shown schematically in Fig.~\ref{fig:schematic},
tracks the motions of monodisperse colloidal spheres
as they are driven back and forth over static potential energy
landscapes created with arrays of 
holographic optical traps.
The samples consist of
colloidal silica spheres $a = 0.75 \pm 0.075~\micron$ in radius
(Bangs Laboratories 5303)
These spheres
were dispersed
 in deionized water and hermetically sealed in a slit pore
formed by bonding the edges of a \#1 cover slip to the face of
a glass microscope slide.  
The glass surfaces were treated
by oxygen plasma etching before assembly to increase their surface
charge and thereby prevent particle deposition.
The sample was rigidly mounted on a Prior Proscan II
translation stage integrated into a Nikon TE2000U
optical microscope, where it was allowed to equilibrate to room
temperature, $T = 296 \pm 2~\unit{K}$.

Previous experimental studies of transport through static light fields
have driven the spheres electrokinetically, optophoretically \cite{lee06a} or
hydrodynamically
\cite{korda02b,macdonald03,ladavac04}.
Others have swept the light field through stationary samples
\cite{koss03,lee05,lee05a,lee06,cizmar06,ricardezvargas06}.
We instead 
used the motorized stage to translate the
entire sample past stationary patterns of optical traps.
All particles consequently traveled past the traps at the same
velocity, $\vec{v}$, without complications due to nonuniform
flow profiles \cite{ladavac04}
and without time-dependent ratchet phenomena \cite{lee06}.
Roughly 5,000 particles were repeatedly passed back and forth
over the same field of view at constant speed and a variety of angles
to build up a statistically well sampled set of data for each
set of conditions.
Repeatedly revisiting the same part of the sample cell with the
same particles minimized effects due to nonuniform
sample thickness and variability in the spheres' properties.

\begin{figure}[!t]
  \centering
  \includegraphics[width=\columnwidth]{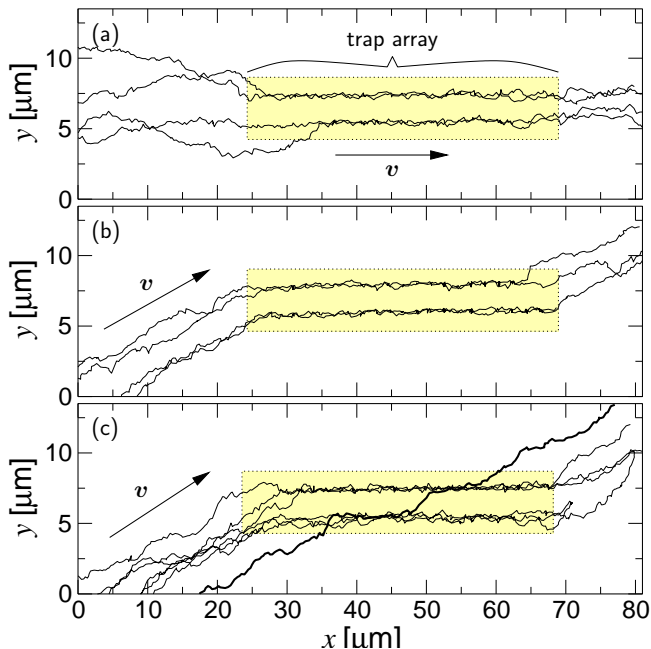}
  \caption{Selected trajectories of $a = 0.75~\micron$
    diameter silica spheres flowing at
    $v = 10~\unit{\micron/s}$ through an array of
    optical traps with $b = 2.025~\micron$ at $P = 0.85~\unit{W}$.
    The shaded boxes indicate the extent of the optical trap array.
    (a) Aligned: $\theta = 0^\circ$.
    (b) Kinetically locked in at $\theta = 10^\circ$.
    (c) Marginally locked in at $\theta_m = 12^\circ$.
    For $\theta \geq \theta_m$, some particles escape 
    the array.}
  \label{fig:trajectories}
\end{figure}

Images of the moving particles were recorded as an uncompressed
digital video stream using an NEC TI-324A video camera and a
Pioneer 520HS digital video recorder.
The combination of a $100\times$ oil immersion objective lens
(Nikon Plan Apo, NA 1.4) and a $0.63\times$ video eyepiece
provides a field of view of $86 \times 63~\unit{\micron^2}$
and a magnification of 0.135~\micron per pixel.
Spheres' images were subsequently
digitized into trajectories with 20~\unit{nm} spatial resolution
at 1/30~\unit{s} intervals using standard methods of digital
video microscopy \cite{crocker96}.  
Typical measured trajectories
are plotted in Fig.~\ref{fig:trajectories}.

The moving spheres encountered 
point-like optical tweezers \cite{ashkin86}
projected in a $20 \times 2$ grid with the holographic
optical trapping technique \cite{dufresne98,curtis02,polin05}.
All of the traps were created from a single beam of laser light at
wavelength $\lambda = 532~\unit{nm}$ and power $P = 0.85~\unit{W}$ 
provided by a frequency-doubled
diode pumped solid state laser (Coherent Verdi).
This beam was diffracted into an array of trap-forming beams
by a phase-only computer-generated
hologram, $\varphi(\vec{\rho})$, an example of which appears in
Fig.~\ref{fig:schematic}(b).
The diffracted beams then were relayed to the objective lens, which
focused them into traps.
Imprinting the hologram on the beam's wavefronts
with a computer-addressed spatial light
modulator (SLM, Hamamatsu X7269 PPM) allows for sequences of
trapping patterns to be projected
with different lattice constants and orientations.
Laser light is reflected into the objective lens with a tuned dichroic
mirror (Chroma Technologies) with a reflectivity of 99.5 percent
at $\lambda = 532~\unit{nm}$.
This mirror transmits light at other wavelengths, which
therefore can be used to create images
of the spheres.

The image of the focused traps in Fig.~\ref{fig:schematic}(c) was
obtained by placing a front surface mirror in the lens' focal
plane.
Enough of the reflected laser light passes through the
dichroic mirror to create a clear low-noise image of the trapping pattern.
Images such as this were used to adaptively improve the traps'
uniformity \cite{polin05}.
After iterative improvement, the traps' intensities typically varied
by less than 15 percent from the mean.

Each colloidal particle experiences an optical trap as radially symmetric
potential energy well whose depth, $V_0(a)$, and width, $\sigma(a)$,
both depend on the particle's shape, size and composition \cite{pelton04a}.
The trap's depth also is proportional
to the total laser power, $P$.
If a particle is deflected enough by its encounter with one trap
to fall into another trap's domain of influence, it can become
kinetically locked in to a commensurate trajectory through
the array of traps \cite{korda02b,ladavac04,pelton04a,gopinathan04}.
If, on the other hand, the driving force is too strong, or the
required deflection angle too steep, the particle
escapes from the traps and runs freely downstream.
In our experiment, the driving force,
$\vec{F}_0(a) = \gamma(a) \, \vec{v}$,
is the hydrodynamic drag on a sphere of radius $a$ driven through
a quiescent fluid at velocity $\vec{v}$.
The viscous drag coefficient, $\gamma(a)$, is proportional to the
fluid's viscosity and accounts for
hydrodynamic coupling to the bounding walls 
\cite{happel91,pozrikidis92,dufresne00,dufresne01}.

The data in Fig.~\ref{fig:trajectories} were obtained
at $v = 10~\unit{\micron/s}$ so that each sphere crossed
the entire field of view within $t = 9~\unit{s}$.
By contrast, the spheres' measured \cite{crocker96,polin05} 
self-diffusion coefficient, $D = 0.10 \pm 0.01~\unit{\micron^2/s}$,
corresponds to a thermally driven displacement of just
$\sqrt{2 D t} = 0.6~\micron$ in the same period.
The associated viscous drag coefficient, 
$\gamma = k_B T / D = 40~\unit{fN~s/\micron}$
suggests that the spheres were driven past the traps
with a maximum force of roughly $F_0 = 0.4~\unit{pN}$.

The distinction between locked-in and free trajectories becomes
clear when the driving force $\vec{F}_0$ is inclined with respect
to the array, as shown in Fig.~\ref{fig:trajectories}(b).
In this case, the locked-in trajectories are deflected by angle
$\theta$ with respect to $\vec{F}_0$.
This deflection is the basis for continuous sorting techniques in
which different fractions of a mixed sample are deflected to
different angles by the same optical intensity field
\cite{korda02b,macdonald03,ladavac04}.

All trajectories become locked in when the driving force is
aligned with the traps, $\theta = 0^\circ$; they all escape
for $\theta = 90^\circ$.
The maximum angle, $\theta_m$, to which a spherical object can be
deflected by a periodic optical intensity field before it escapes
has been predicted
to depend exceptionally strongly on the object's radius
\cite{ladavac04,pelton04a,sancho05}.
This is a purely kinematic effect, and not the result
of thermal activation over potential energy barriers, as
has been suggested \cite{macdonald03}.

Modeling the optical traps as Gaussian potential energy
wells separated by distance $b$ yields \cite{ladavac04,pelton04a} 
\begin{equation}
  \label{eq:thetam}
  \sin \theta_m = \abs{S(a)} \,
  \exp\left(-\frac{b^2}{8 \sigma^2(a)}\right),
\end{equation}
with the prefactor,
\begin{equation}
  S(a) = \frac{2}{\sqrt{e}} \, \frac{V_0(a)}{\sigma(a) F_0(a)},
  \label{eq:s}
\end{equation}
reflecting a balance between trapping and driving forces at
the point of escape.
The traps' effective width is
given by
\begin{equation}
  \sigma^2(a) = a^2 + \frac{\lambda^2}{4n^2},
  \label{eq:sigma}
\end{equation}
where $n$ is the fluid's index of refraction.
For the particles in our study, dispersed in water with
$n = 1.33$, $\sigma = 0.85~\micron$.
Similar results are obtained for more general periodic landscapes
\cite{pelton04a}, with Eq.~(\ref{eq:thetam}) representing the
leading term in a Fourier expansion of the trapping potential.
Equation~(\ref{eq:thetam}) also applies for arrays of barriers rather
than wells \cite{pelton04a}.

\begin{figure}[b!]
  \centering
  \includegraphics[width=\columnwidth]{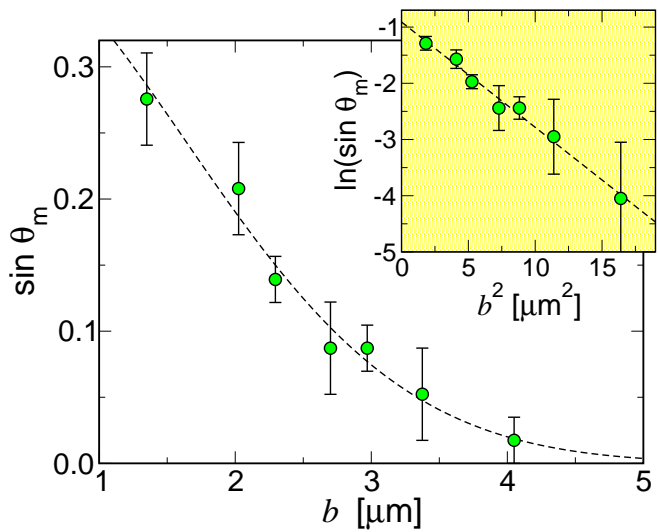}
  \caption{Dependence of the marginally locked in angle as a 
    function of lattice constant.  
    Inset: The data replotted to emphasize the comparison to 
    Eq.~(\protect\ref{eq:thetam}).
  }
  \label{fig:lockin}
\end{figure}

According to Eq.~(\ref{eq:thetam}),
the marginally locked-in angle, $\theta_m$,
depends very strongly on the ratio $b/\sigma(a)$, and
thus on particle size for
$a > \lambda$.
This is the basis for the assertion \cite{ladavac04} that sorting
by transport through a periodic landscape can offer
exponential size selectivity.
This prediction, however, results from limiting arguments
that have not been tested experimentally.
More rigorous results are available only in the limit
that $V_0 \ll F_0 \sigma \approx k_B T$, in which case
thermal forces cannot be neglected and the escape
transition is expected to be less dramatic \cite{gleeson06}.

We explicitly tested the predicted dependence on trap
separation, $b$, and laser power by driving monodisperse
particles past adaptively optimized trap arrays at fixed speed, 
$v$, over a range of angles, gauging the marginal angle
by the suddenly increasing proportion of escaping trajectories.
As $\theta$ approaches $\theta_m$, particles' trajectories
become increasingly sensitive to variations in the traps'
intensities, to thermal fluctuations, and to small differences
in individual particles' radii.
A typical escape transition is apparent in the selected trajectories
plotted in Fig.~\ref{fig:trajectories}(c).
It is reasonable to expect that the first observed escape events
would involve the smallest particles interacting with the weakest traps.
Analyzing particle tracks, however, did not provide sufficient
size resolution to test this directly.
The marginally locked-in angle for a particular choice of $b$
was determined by analyzing
roughly 2000 such trajectories for each value of $\theta$ 
ranging from $0^\circ$ to $30^\circ$ in $2^\circ$ increments.

Results for $\theta_m(b)$, 
obtained at fixed speed, $v$, and laser power, $P$,
are plotted in Fig.~\ref{fig:lockin}.
The dashed curve is a one-parameter fit to Eq.~(\ref{eq:thetam}),
where only the overall scale, $S(a)$, is treated as a free parameter.
The fit value, $S(a) = 0.40 \pm 0.01$ corresponds to a well depth
of roughly $100~k_B T$, which is reasonable for traps powered with
2~\unit{mW} of light.
The experimental results' excellent agreement with the prediction
suggests that Eq.~(\ref{eq:thetam}) quantitatively describes
colloidal transport through an array of optical traps.

\begin{figure}[!t]
  \centering
  \includegraphics[width=\columnwidth]{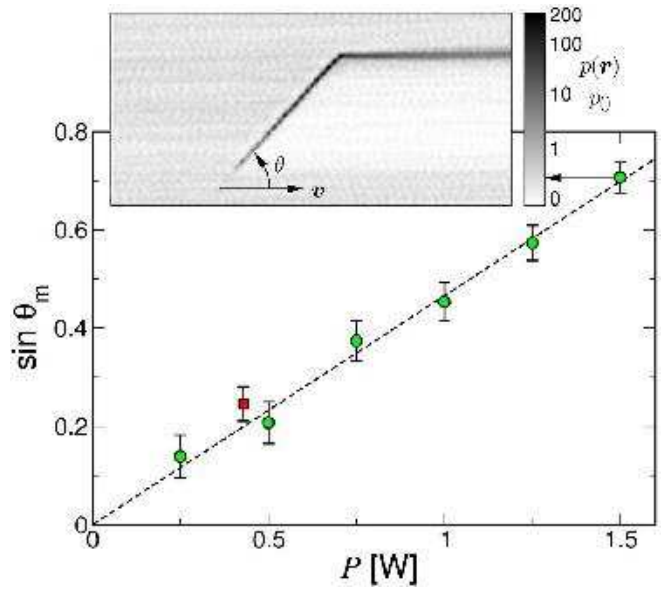}  
  \caption{Dependence of the marginally locked in angle on total 
    projected laser power for fixed values of lattice constant,
    $b = 1.35~\micron$, and speed,
    $v = 46~\unit{\micron/s}$.
    The square point is replotted from Fig.~\protect\ref{fig:lockin}.
    Inset: Probability density $p(\vec{r})$ to
    find a particle at $\vec{r}$, normalized by the bulk value,
    $p_0$, at $\theta = 45^\circ$.  The downstream
    shadow in $p(\vec{r})$ demonstrates effectively perfect
    large-angle deflection for $P = 1.5~\unit{W}$.
    Average over 500 trajectories.}
  \label{fig:power}
\end{figure}

The marginally locked-in angle can be 
tuned for maximum sensitivity by
adjusting $b$.  
Its magnitude can be set independently through the prefactor, $S(a)$.
In particular, $S(a)$ should scale linearly with
laser power, $P$, through its dependence on $V_0$.
Thus $\theta_m$ for a given size of particle
can be adjusted with laser power, for fixed $v$, up to the point that
particles begin to get stuck in the traps.
The data in Fig.~\ref{fig:power} demonstrate that deflection angles
as large as $\theta_m = 45^\circ$ can be attained in this way.
They therefore contradict the assertion \cite{macdonald03} that 
arrays of discrete traps are incapable of deflecting trajectories
over large angles.

The data in Fig.~\ref{fig:power} were obtained with a single row of 20 traps
at fixed inter-trap separation of
$b = 1.35~\micron$, and with particles driven at $v = 46~\unit{\micron/s}$.
Using a single line of traps increases the possibility that particles
might leak through the array, but doubles the accessible range of
laser powers.
Data from Fig.~\ref{fig:lockin} at the same particle speed and lattice constant fall on
the same curve once the laser power is rescaled.
The observed linear dependence of $\sin \theta_m$ on laser power, $P$,
confirms the predicted dependence on the prefactor,
$S(a)$ in Eq.~(\ref{eq:thetam}).

The plot inset into Fig.~\ref{fig:power} provides an overview
of the data
obtained at $\theta_m = 45^\circ$ and $P = 1.5~\unit{W}$.
It shows the probability
distribution $p(\vec{r})$ for finding a particle within
200~\unit{nm} of $\vec{r}$, integrated over 500 trajectories,
comprising roughly 30,000 separate particle images.
The probability distribution is normalized by the probability,
$p_0$, of finding a particle in the undeflected stream over the
same period.  Darker regions indicate a higher-than-average
probability density and lighter regions indicate lower-than-average
probabilities.
Details in the distribution are emphasized with
a nonlinear color table, which also is inset.
The results show that 
particles are strongly concentrated in the traps themselves
and work their way up the array until they escape at its end.
The nearly perfect deflection of the incident stream of 
particles leaves a sharply
defined shadow in the probability density downstream of the 
traps.  A total of 15 of the 500 particles escaped the array near
its end, most likely representing the smallest end of the
particle size distribution.

Confirming the marginally locked-in angle's dependence 
on trap separation and laser power supports the assumptions made in 
deriving Eq.~(\ref{eq:thetam}), particularly because the
predicted form for $\sigma(a)$ agrees quantitatively with
experimental results.
This provides additional, albeit indirect, support
for the prediction that athermal sieving by periodically modulated
landscapes can sort objects with exponential size selectivity.
It leaves open questions regarding the nature of transitions among
different commensurate locked-in states in more extensive
two-dimensional lattices.
It also does not address the nature of colloidal transport through
aperiodic landscapes, such as quasiperiodic arrays of optical traps.
Experiments to address these questions are in progress.

This work was supported by the National Science Foundation
through Grant Number DMR-0451589.


\begin{thebibliography}{29}
\expandafter\ifx\csname natexlab\endcsname\relax\def\natexlab#1{#1}\fi
\expandafter\ifx\csname bibnamefont\endcsname\relax
  \def\bibnamefont#1{#1}\fi
\expandafter\ifx\csname bibfnamefont\endcsname\relax
  \def\bibfnamefont#1{#1}\fi
\expandafter\ifx\csname citenamefont\endcsname\relax
  \def\citenamefont#1{#1}\fi
\expandafter\ifx\csname url\endcsname\relax
  \def\url#1{\texttt{#1}}\fi
\expandafter\ifx\csname urlprefix\endcsname\relax\def\urlprefix{URL }\fi
\providecommand{\bibinfo}[2]{#2}
\providecommand{\eprint}[2][]{\url{#2}}

\bibitem[{\citenamefont{Carlson et~al.}(1997)\citenamefont{Carlson, Gabel,
  Chan, and Austin}}]{carlson97}
\bibinfo{author}{\bibfnamefont{R.~H.}~\bibnamefont{Carlson}}, 
	\bibinfo{author}{\bibfnamefont{C.~V.}
  \bibnamefont{Gabel}}, \bibinfo{author}{\bibfnamefont{S.~S.}
  \bibnamefont{Chan}}, \bibnamefont{and} \bibinfo{author}{\bibfnamefont{R.~H.}
  \bibnamefont{Austin}}, \bibinfo{journal}{Phys. Rev. Lett.}
  \textbf{\bibinfo{volume}{79}}, \bibinfo{pages}{2149} (\bibinfo{year}{1997}).

\bibitem[{\citenamefont{Duke and Austin}(1997)}]{duke97}
\bibinfo{author}{\bibfnamefont{T.~A.~J.} \bibnamefont{Duke}} \bibnamefont{and}
  \bibinfo{author}{\bibfnamefont{R.~H.} \bibnamefont{Austin}},
  \bibinfo{journal}{Phys. Rev. Lett.} \textbf{\bibinfo{volume}{80}},
  \bibinfo{pages}{1552} (\bibinfo{year}{1997}).

\bibitem[{\citenamefont{Chou et~al.}(1999)\citenamefont{Chou, Bakajin, Turner,
  Duke, Chan, Cox, Craighead, and Austin}}]{chou99}
\bibinfo{author}{\bibfnamefont{C.-F.} \bibnamefont{Chou}},
  \bibinfo{author}{\bibfnamefont{O.}~\bibnamefont{Bakajin}},
  \bibinfo{author}{\bibfnamefont{S.~W.~P.} \bibnamefont{Turner}},
  \bibinfo{author}{\bibfnamefont{T.~A.~J.} \bibnamefont{Duke}},
  \bibinfo{author}{\bibfnamefont{S.~S.} \bibnamefont{Chan}},
  \bibinfo{author}{\bibfnamefont{E.~C.} \bibnamefont{Cox}},
  \bibinfo{author}{\bibfnamefont{H.~G.} \bibnamefont{Craighead}},
  \bibnamefont{and} \bibinfo{author}{\bibfnamefont{R.~H.}
  \bibnamefont{Austin}}, \bibinfo{journal}{Proc. Nat. Acad. Sci.}
  \textbf{\bibinfo{volume}{96}}, \bibinfo{pages}{13762} (\bibinfo{year}{1999}).

\bibitem[{\citenamefont{Chou et~al.}(2000)\citenamefont{Chou, Austin, Bakajin,
  Tegenfeldt, Castelino, Chan, Cox, Craighead, Darnton, Duke et~al.}}]{chou00}
\bibinfo{author}{\bibfnamefont{C.~F.} \bibnamefont{Chou}},
  \bibinfo{author}{\bibfnamefont{R.~H.} \bibnamefont{Austin}},
  \bibinfo{author}{\bibfnamefont{O.}~\bibnamefont{Bakajin}},
  \bibinfo{author}{\bibfnamefont{J.~O.} \bibnamefont{Tegenfeldt}},
  \bibinfo{author}{\bibfnamefont{J.~A.} \bibnamefont{Castelino}},
  \bibinfo{author}{\bibfnamefont{S.~S.} \bibnamefont{Chan}},
  \bibinfo{author}{\bibfnamefont{E.~C.} \bibnamefont{Cox}},
  \bibinfo{author}{\bibfnamefont{H.}~\bibnamefont{Craighead}},
  \bibinfo{author}{\bibfnamefont{N.}~\bibnamefont{Darnton}},
  \bibinfo{author}{\bibfnamefont{T.}~\bibnamefont{Duke}}, \bibnamefont{et~al.},
  \bibinfo{journal}{Electrophoresis} \textbf{\bibinfo{volume}{21}},
  \bibinfo{pages}{81} (\bibinfo{year}{2000}).

\bibitem[{\citenamefont{Hunt et~al.}(2004)\citenamefont{Hunt, Lee, and
  Westervelt}}]{hunt04}
\bibinfo{author}{\bibfnamefont{T.~P.} \bibnamefont{Hunt}},
  \bibinfo{author}{\bibfnamefont{H.}~\bibnamefont{Lee}}, \bibnamefont{and}
  \bibinfo{author}{\bibfnamefont{R.~M.} \bibnamefont{Westervelt}},
  \bibinfo{journal}{Appl. Phys. Lett.} \textbf{\bibinfo{volume}{85}},
  \bibinfo{pages}{6421} (\bibinfo{year}{2004}).

\bibitem[{\citenamefont{Chiou et~al.}(2005)\citenamefont{Chiou, Ohta, and
  Wu}}]{chiou05}
\bibinfo{author}{\bibfnamefont{P.~Y.} \bibnamefont{Chiou}},
  \bibinfo{author}{\bibfnamefont{T.}~\bibnamefont{Ohta}, \bibfnamefont{Aaron}},
  \bibnamefont{and} \bibinfo{author}{\bibfnamefont{M.~C.} \bibnamefont{Wu}},
  \bibinfo{journal}{Nature} \textbf{\bibinfo{volume}{436}},
  \bibinfo{pages}{370} (\bibinfo{year}{2005}).

\bibitem[{\citenamefont{Korda et~al.}(2002)\citenamefont{Korda, Taylor, and
  Grier}}]{korda02b}
\bibinfo{author}{\bibfnamefont{P.~T.} \bibnamefont{Korda}},
  \bibinfo{author}{\bibfnamefont{M.~B.} \bibnamefont{Taylor}},
  \bibnamefont{and} \bibinfo{author}{\bibfnamefont{D.~G.} \bibnamefont{Grier}},
  \bibinfo{journal}{Phys. Rev. Lett.} \textbf{\bibinfo{volume}{89}},
  \bibinfo{pages}{128301} (\bibinfo{year}{2002}).

\bibitem[{\citenamefont{MacDonald et~al.}(2003)\citenamefont{MacDonald,
  Spalding, and Dholakia}}]{macdonald03}
\bibinfo{author}{\bibfnamefont{M.~P.} \bibnamefont{MacDonald}},
  \bibinfo{author}{\bibfnamefont{G.~C.} \bibnamefont{Spalding}},
  \bibnamefont{and} \bibinfo{author}{\bibfnamefont{K.}~\bibnamefont{Dholakia}},
  \bibinfo{journal}{Nature} \textbf{\bibinfo{volume}{426}},
  \bibinfo{pages}{421} (\bibinfo{year}{2003}).

\bibitem[{\citenamefont{Ladavac et~al.}(2004)\citenamefont{Ladavac, Kasza, and
  Grier}}]{ladavac04}
\bibinfo{author}{\bibfnamefont{K.}~\bibnamefont{Ladavac}},
  \bibinfo{author}{\bibfnamefont{K.}~\bibnamefont{Kasza}}, \bibnamefont{and}
  \bibinfo{author}{\bibfnamefont{D.~G.} \bibnamefont{Grier}},
  \bibinfo{journal}{Phys. Rev. E} \textbf{\bibinfo{volume}{70}},
  \bibinfo{pages}{010901(R)} (\bibinfo{year}{2004}).

\bibitem[{\citenamefont{Pelton et~al.}(2004)\citenamefont{Pelton, Ladavac, and
  Grier}}]{pelton04a}
\bibinfo{author}{\bibfnamefont{M.}~\bibnamefont{Pelton}},
  \bibinfo{author}{\bibfnamefont{K.}~\bibnamefont{Ladavac}}, \bibnamefont{and}
  \bibinfo{author}{\bibfnamefont{D.~G.} \bibnamefont{Grier}},
  \bibinfo{journal}{Phys. Rev. E} \textbf{\bibinfo{volume}{70}},
  \bibinfo{pages}{031108} (\bibinfo{year}{2004}).

\bibitem[{\citenamefont{Lee and Grier}(2006{\natexlab{a}})}]{lee06a}
\bibinfo{author}{\bibfnamefont{S.-H.} \bibnamefont{Lee}} \bibnamefont{and}
  \bibinfo{author}{\bibfnamefont{D.~G.} \bibnamefont{Grier}},
  \bibinfo{journal}{Phys. Rev. Lett.} \textbf{\bibinfo{volume}{96}},
  \bibinfo{pages}{190601} (\bibinfo{year}{2006}{\natexlab{a}}).

\bibitem[{\citenamefont{Lee et~al.}(2005)\citenamefont{Lee, Ladavac, Polin, and
  Grier}}]{lee05}
\bibinfo{author}{\bibfnamefont{S.-H.} \bibnamefont{Lee}},
  \bibinfo{author}{\bibfnamefont{K.}~\bibnamefont{Ladavac}},
  \bibinfo{author}{\bibfnamefont{M.}~\bibnamefont{Polin}}, \bibnamefont{and}
  \bibinfo{author}{\bibfnamefont{D.~G.} \bibnamefont{Grier}},
  \bibinfo{journal}{Phys. Rev. Lett.} \textbf{\bibinfo{volume}{94}},
  \bibinfo{pages}{110601} (\bibinfo{year}{2005}).

\bibitem[{\citenamefont{Lee and Grier}(2005)}]{lee05a}
\bibinfo{author}{\bibfnamefont{S.-H.} \bibnamefont{Lee}} \bibnamefont{and}
  \bibinfo{author}{\bibfnamefont{D.~G.} \bibnamefont{Grier}},
  \bibinfo{journal}{Phys. Rev. E} \textbf{\bibinfo{volume}{71}},
  \bibinfo{pages}{060102(R)} (\bibinfo{year}{2005}).

\bibitem[{\citenamefont{Lee and Grier}(2006{\natexlab{b}})}]{lee06}
\bibinfo{author}{\bibfnamefont{S.-H.} \bibnamefont{Lee}} \bibnamefont{and}
  \bibinfo{author}{\bibfnamefont{D.~G.} \bibnamefont{Grier}},
  \bibinfo{journal}{J. Phys.: Condens. Matt.} \textbf{\bibinfo{volume}{17}},
  \bibinfo{pages}{S3685} (\bibinfo{year}{2006}{\natexlab{b}}).

\bibitem[{\citenamefont{\u{C}i\u{z}m\'{a}r
  et~al.}(2006)\citenamefont{\u{C}i\u{z}m\'{a}r, \u{S}iler, \u{S}er\'{y},
  Zem\'{a}nek, Garc\'es-Ch\'avez, and Dholakia}}]{cizmar06}
\bibinfo{author}{\bibfnamefont{T.}~\bibnamefont{\u{C}i\u{z}m\'{a}r}},
  \bibinfo{author}{\bibfnamefont{M.}~\bibnamefont{\u{S}iler}},
  \bibinfo{author}{\bibfnamefont{M.}~\bibnamefont{\u{S}er\'{y}}},
  \bibinfo{author}{\bibfnamefont{P.}~\bibnamefont{Zem\'{a}nek}},
  \bibinfo{author}{\bibfnamefont{V.}~\bibnamefont{Garc\'es-Ch\'avez}},
  \bibnamefont{and} \bibinfo{author}{\bibfnamefont{K.}~\bibnamefont{Dholakia}},
  \bibinfo{journal}{Phys. Rev. B} \textbf{\bibinfo{volume}{74}},
  \bibinfo{pages}{035105} (\bibinfo{year}{2006}).

\bibitem[{\citenamefont{Koss and Grier}(2003)}]{koss03}
\bibinfo{author}{\bibfnamefont{B.~A.} \bibnamefont{Koss}} \bibnamefont{and}
  \bibinfo{author}{\bibfnamefont{D.~G.} \bibnamefont{Grier}},
  \bibinfo{journal}{Appl. Phys. Lett.} \textbf{\bibinfo{volume}{82}},
  \bibinfo{pages}{3985} (\bibinfo{year}{2003}).

\bibitem[{\citenamefont{Ric\'ardez-Vargas
  et~al.}(2006)\citenamefont{Ric\'ardez-Vargas, Rodr\'{\i}guez-Montero,
  Romos-Garc\'{\i}a, and Volke-Sup\'ulveda}}]{ricardezvargas06}
\bibinfo{author}{\bibfnamefont{I.}~\bibnamefont{Ric\'ardez-Vargas}},
  \bibinfo{author}{\bibfnamefont{P.}~\bibnamefont{Rodr\'{\i}guez-Montero}},
  \bibinfo{author}{\bibfnamefont{R.}~\bibnamefont{Romos-Garc\'{\i}a}},
  \bibnamefont{and}
  \bibinfo{author}{\bibfnamefont{K.}~\bibnamefont{Volke-Sup\'ulveda}},
  \bibinfo{journal}{Appl. Phys. Lett.} \textbf{\bibinfo{volume}{88}},
  \bibinfo{pages}{121116} (\bibinfo{year}{2006}).

\bibitem[{\citenamefont{Crocker and Grier}(1996)}]{crocker96}
\bibinfo{author}{\bibfnamefont{J.~C.} \bibnamefont{Crocker}} \bibnamefont{and}
  \bibinfo{author}{\bibfnamefont{D.~G.} \bibnamefont{Grier}},
  \bibinfo{journal}{J. Colloid Interface Sci.} \textbf{\bibinfo{volume}{179}},
  \bibinfo{pages}{298} (\bibinfo{year}{1996}).

\bibitem[{\citenamefont{Ashkin et~al.}(1986)\citenamefont{Ashkin, Dziedzic,
  Bjorkholm, and Chu}}]{ashkin86}
\bibinfo{author}{\bibfnamefont{A.}~\bibnamefont{Ashkin}},
  \bibinfo{author}{\bibfnamefont{J.~M.} \bibnamefont{Dziedzic}},
  \bibinfo{author}{\bibfnamefont{J.~E.} \bibnamefont{Bjorkholm}},
  \bibnamefont{and} \bibinfo{author}{\bibfnamefont{S.}~\bibnamefont{Chu}},
  \bibinfo{journal}{Opt. Lett.} \textbf{\bibinfo{volume}{11}},
  \bibinfo{pages}{288} (\bibinfo{year}{1986}).

\bibitem[{\citenamefont{Dufresne and Grier}(1998)}]{dufresne98}
\bibinfo{author}{\bibfnamefont{E.~R.} \bibnamefont{Dufresne}} \bibnamefont{and}
  \bibinfo{author}{\bibfnamefont{D.~G.} \bibnamefont{Grier}},
  \bibinfo{journal}{Rev. Sci. Instr.} \textbf{\bibinfo{volume}{69}},
  \bibinfo{pages}{1974} (\bibinfo{year}{1998}).

\bibitem[{\citenamefont{Curtis et~al.}(2002)\citenamefont{Curtis, Koss, and
  Grier}}]{curtis02}
\bibinfo{author}{\bibfnamefont{J.~E.} \bibnamefont{Curtis}},
  \bibinfo{author}{\bibfnamefont{B.~A.} \bibnamefont{Koss}}, \bibnamefont{and}
  \bibinfo{author}{\bibfnamefont{D.~G.} \bibnamefont{Grier}},
  \bibinfo{journal}{Opt. Comm.} \textbf{\bibinfo{volume}{207}},
  \bibinfo{pages}{169} (\bibinfo{year}{2002}).

\bibitem[{\citenamefont{Polin et~al.}(2005)\citenamefont{Polin, Ladavac, Lee,
  Roichman, and Grier}}]{polin05}
\bibinfo{author}{\bibfnamefont{M.}~\bibnamefont{Polin}},
  \bibinfo{author}{\bibfnamefont{K.}~\bibnamefont{Ladavac}},
  \bibinfo{author}{\bibfnamefont{S.-H.} \bibnamefont{Lee}},
  \bibinfo{author}{\bibfnamefont{Y.}~\bibnamefont{Roichman}}, \bibnamefont{and}
  \bibinfo{author}{\bibfnamefont{D.~G.} \bibnamefont{Grier}},
  \bibinfo{journal}{Opt. Express} \textbf{\bibinfo{volume}{13}},
  \bibinfo{pages}{5831} (\bibinfo{year}{2005}).

\bibitem[{\citenamefont{Gopinathan and Grier}(2004)}]{gopinathan04}
\bibinfo{author}{\bibfnamefont{A.}~\bibnamefont{Gopinathan}} \bibnamefont{and}
  \bibinfo{author}{\bibfnamefont{D.~G.} \bibnamefont{Grier}},
  \bibinfo{journal}{Phys. Rev. Lett.} \textbf{\bibinfo{volume}{92}},
  \bibinfo{pages}{130602} (\bibinfo{year}{2004}).

\bibitem[{\citenamefont{Happel and Brenner}(1991)}]{happel91}
\bibinfo{author}{\bibfnamefont{J.}~\bibnamefont{Happel}} \bibnamefont{and}
  \bibinfo{author}{\bibfnamefont{H.}~\bibnamefont{Brenner}},
  \emph{\bibinfo{title}{Low Reynolds Number Hydrodynamics}}
  (\bibinfo{publisher}{Kluwer}, \bibinfo{address}{Dordrecht},
  \bibinfo{year}{1991}).

\bibitem[{\citenamefont{Pozrikidis}(1992)}]{pozrikidis92}
\bibinfo{author}{\bibfnamefont{C.}~\bibnamefont{Pozrikidis}},
  \emph{\bibinfo{title}{Boundary Integral and Singularity Methods for
  Linearized Viscous Flow}} (\bibinfo{publisher}{Cambridge University Press},
  \bibinfo{address}{New York}, \bibinfo{year}{1992}).

\bibitem[{\citenamefont{Dufresne et~al.}(2000)\citenamefont{Dufresne, Squires,
  Brenner, and Grier}}]{dufresne00}
\bibinfo{author}{\bibfnamefont{E.~R.} \bibnamefont{Dufresne}},
  \bibinfo{author}{\bibfnamefont{T.~M.} \bibnamefont{Squires}},
  \bibinfo{author}{\bibfnamefont{M.~P.} \bibnamefont{Brenner}},
  \bibnamefont{and} \bibinfo{author}{\bibfnamefont{D.~G.} \bibnamefont{Grier}},
  \bibinfo{journal}{Phys. Rev. Lett.} \textbf{\bibinfo{volume}{85}},
  \bibinfo{pages}{3317} (\bibinfo{year}{2000}).

\bibitem[{\citenamefont{Dufresne et~al.}(2001)\citenamefont{Dufresne, Altman,
  and Grier}}]{dufresne01}
\bibinfo{author}{\bibfnamefont{E.~R.} \bibnamefont{Dufresne}},
  \bibinfo{author}{\bibfnamefont{D.}~\bibnamefont{Altman}}, \bibnamefont{and}
  \bibinfo{author}{\bibfnamefont{D.~G.} \bibnamefont{Grier}},
  \bibinfo{journal}{Europhys. Lett.} \textbf{\bibinfo{volume}{53}},
  \bibinfo{pages}{264} (\bibinfo{year}{2001}).

\bibitem[{\citenamefont{Sancho et~al.}(2005)\citenamefont{Sancho, Khoury,
  Lindenberg, and Lacasta}}]{sancho05}
\bibinfo{author}{\bibfnamefont{J.~M.} \bibnamefont{Sancho}},
  \bibinfo{author}{\bibfnamefont{M.}~\bibnamefont{Khoury}},
  \bibinfo{author}{\bibfnamefont{K.}~\bibnamefont{Lindenberg}},
  \bibnamefont{and} \bibinfo{author}{\bibfnamefont{A.~M.}
  \bibnamefont{Lacasta}}, \bibinfo{journal}{J. Phys.: Condens. Matt.}
  \textbf{\bibinfo{volume}{17}}, \bibinfo{pages}{S4151} (\bibinfo{year}{2005}).

\bibitem[{\citenamefont{Gleeson et~al.}(2006)\citenamefont{Gleeson, Sancho,
  Lacasta, and Lindenberg}}]{gleeson06}
\bibinfo{author}{\bibfnamefont{J.~P.} \bibnamefont{Gleeson}},
  \bibinfo{author}{\bibfnamefont{J.~M.} \bibnamefont{Sancho}},
  \bibinfo{author}{\bibfnamefont{A.~M.} \bibnamefont{Lacasta}},
  \bibnamefont{and}
  \bibinfo{author}{\bibfnamefont{K.}~\bibnamefont{Lindenberg}},
  \bibinfo{journal}{Phys. Rev. E} \textbf{\bibinfo{volume}{73}},
  \bibinfo{pages}{041102} (\bibinfo{year}{2006}).

\end{thebibliography}

\end{document}